\documentclass[11pt]{article}
\usepackage{latexsym}
\usepackage{amsthm}
\usepackage{mathrsfs}
\usepackage{amsfonts}
\usepackage{enumerate}
\usepackage{amsmath}
\usepackage{amssymb}
\begin{document}
\begin{center}
\null\vspace{2cm}
{\large {\bf Hawking non-thermal and Purely thermal radiations of Kerr-de Sitter black hole by Hamilton-Jacobi method}}\\
\vspace{2cm}
M. Ilias Hossain\footnote{E-mail: $ilias_-math@yahoo.com$}\\
{\it Department of Mathematics, Rajshahi University,  Rajshahi - 6205, Bangladesh}\\
M. Atiqur Rahman\footnote{E-mail: $atirubd@yahoo.com$}\\
{\it Department of Applied Mathematics, Rajshahi University, Rajshahi - 6205, Bangladesh}\\

\end{center}
\vspace{3cm}
\centerline{\bf Abstract}
\baselineskip=18pt
\bigskip
Incorporating Parikh and Wilczek\rq's opinion to the  Kerr de-Sitter (KdS) black hole Hawking non-thermal and purely thermal radiations have been investigated  using Hamilton-Jacobi  method. We have taken the background spacetime of KdS black hole as dynamical, involving the self-gravitation effect of the emitted particles, energy and angular momentum has been taken as conserved and show that the tunneling rate is related to the change of Bekenstein-Hawking entropy and the derived emission spectrum deviates from the pure thermal spectrum. The explored results gives a correction to the Hawking radiation of KdS black hole.
\vspace{0.5cm}\\
{\bf Keywords: Massive Particle Tunneling, KdS black hole.}\\
\vfill

\newpage

\section{Introduction}\label{sec1}
A prominent role in black hole physics, the discovery by Hawking in $1974$ that black hole can emit particles \cite{one,two} from the event horizon. At once time, the picture of the primary state of composing matter leads to a paradoxical claim of black hole information loss. This is so-called  ``the paradox of black hole information loss". Taking quantum process into account, the black holes mass becomes small and smaller and eventually completely evaporated \cite{three} that is the situation is changed.  In the tunneling mechanism, virtual particles and anti-particles create due to vacuum fluctuation and a positive energy particle tunnels out of the horizon and materializes as a real particle, while the negative energy particle tunnels into the horizon and is absorbed and moved to infinite distance whenever the positive energy particle is left outside the horizon  and forms the Hawking radiation.  A semiclassical tunneling method innovated by Kraus and Wilczek \cite{four, five} and then reinterpreted  by Parikh and Wilczek \cite{six, seven} and have shown that the actual emission spectrum deviates from the purely thermal. People have growing interest and therefore several works have been done for further development of the tunneling approach, but all of them are only focused on Hawking radiation of various black hole spacetime such as those in de Sitter \cite{seven,eight, nine, ten, eleven}, anti-de Sitter \cite{twelve, thirteen, fourteen} spacetimes, charged black holes \cite{ten, eleven,fifteen}, rotating black holes \cite{sixteen} and many other cases in references \cite{seventeen,eighteen,nineteen,twenty,twenty one,twenty two,twenty three,twenty four,twenty five,twenty six,twenty seven,twenty eight,twenty nine,thirty,thirty one, thirty two, thirty three,thirty four,thirty five,thirty six}. All of these works are limited to massless particle and gives a correction to the emission rate arising from loss of mass of the black hole corresponding to the energy carried by radiated quantum when the energy conservation and the self gravitation interaction are considered . To calculate the action, most of the researchers introduced Painlev\'e or dragging coordinates that are well regular at the horizon of the black hole and found out the motion equation of the particle, and then calculated Hamilton equation to get it but we avoid these type of coordinates and in this method the tunneling rate is  related to the change of Bekenstein-Hawking entropy \cite{six, seven}.

Based on semiclassical tunneling picture, Angheben et al. \cite{thirty seven} proposed `Hamilton-Jacobi method' and in fact this method is an extension of the complex path analysis proposed by Padmanabhan et al. \cite{thirty eight,thirty nine,fourty}. This method involves calculating the imaginary part of the action from relativistic Hamilton-Jacobi equation in which the derive radiation spectrum was only a leading term due to the fact that the self-gravitation interaction and energy conservation of emitted particle were ignored. Parikh and Wilczek \cite{six} have shown that  the true radiation spectrum is not strictly thermal but satisfies the underlying unitary theory when self-gravitation interaction and energy conservation are considered and the background geometry of a radiating black hole can be altered (unfixed) with the loss of radiated energy.

Kerner and Mann promoted quantum tunneling methods for various black hole spacetime \cite{fourty one} and considering this process Chen, Zu and Yang \cite{fourty two} reformed Hamilton-Jacobi  method for massive particle tunneling and investigate  Hawking radiation of Kerr-NUT  black hole \cite{fourty three}, the charged black hole with a global monopole \cite{fourty four} and also applied to higher dimensional black holes \cite{fourty five}, black holes in string theory \cite{fourty six} , black strings \cite{fourty seven}, accelerating and rotaing black holes \cite{fourty eight} and many other black holes in references \cite{fourty nine, fifty}. Following their work, several researches have been carried out as charged particle tunneling \cite{ten, sixteen, fifty one, fifty two} and all the results supported Parikh and Wilczek\rq s \cite{six} opinion and gave a correction to the Hawking pure thermal spectrum.

In recent times, we have   developed Hamilton-Jacobi method and investigated the Hawking purely thermal and non-thermal radiations of the SdS \cite{fifty three} black hole where the position of the black hole horizon is taken in an infinite series of black hole\rq s parameters which is the new innovation in our research as well as the new line element is derived by using near horizon approximation of spacetime  so that the spacetime metric becomes dynamical and self-gravitation interaction  are taken into account and the changed of background geometry can be treated as the loss of radiated energy of the black hole. In this paper, we have been applied the same method to investigate the Hawking radiation of KdS black hole. The method of Chen et al. \cite{fourty two} is used  to describe Hawking radiation from the action of radiation particles. Since our prime concern of this work is to calculate the imaginary part of action from Hamilton-Jacobi equation avoid by exploring the equation of motion of the radiation particle in Painlev\'e coordinate system and calculating the Hamilton equation. The equation of motion of massive particles are different from massless particle though the radiation particles do not vary. After considering the self-gravitational interaction and the unfixed background spacetime, the derived radiation spectrum deviates from the purely thermal one and the tunneling rate is related to the change of Bekenstein-Hawking entropy. It is noticed that the cosmological constant plays an important role in our research because the accelerating expansion of our universe indicates the cosmological constant might be a positive one \cite{fifty four, fifty five, fifty six}, and the conjecture about de Sitter/conformal field theory (CFT) correspondence \cite{fifty seven,fifty eight} has been suggested that there is a dual relation between  quantum gravity on a dS space and Euclidean conformal field theory (CFT) on a boundary of dS space \cite{fifty nine}. The outgoing particles tunnel from black hole horizon and incoming particles tunnel from cosmological horizon and formed Hawking radiation and the incoming particles can fall into the horizon along classically permitted trajectories for black hole horizon, but outgoing particles can fall classically out of the horizon for cosmological horizon. Thus our study of black hole on Kerr-de Sitter black hole is of great consequence and significant.

The outline of the  article is the following: The later section describes the KdS black hole spacetime with the position of event horizon and for the convenient of discussion of the Hawking radiation, the new line element of KdS black hole is derived here near the event horizon in section 3. The unfixed background spacetime and the self-gravitational interaction are taken into account, we also review the Hawking radiation of KdS black hole from massive particle tunneling method in section 3. In section 4, we have developed the Hawking purely thermal rate from non-thermal rate. Finally, section 5 encloses our remarks.

\section{Kerr-de Sitter black hole}\label{sec2}

The line element, describing Kerr-de Sitter solution with cosmological constant $\Lambda(=3/\ell^2)$, rotating black hole
in four dimensional spacetime with asymptotic-de Sitter behavior in the Boyer-Lindguist coordinates \cite{sixty} is given by
\begin{eqnarray}
ds^2=-\frac{f(r)-f(\theta)a^2\sin^2\theta}{\rho^2}dt^2+\frac{\rho^2}{f(r)}dr^2+\frac{\rho^2}{f(\theta)}d\theta^2+\frac{f(\theta)(r^2+a^2)^2-f(r)a^2\sin^2\theta}{\rho^2 \Sigma^2}\sin^2\theta d\phi^2\nonumber\\
-\frac{2a[(r^2+a^2)f(\theta)-f(r)]\sin^2\theta}{\rho^2 \Sigma}dt d\phi,\label{eq1}
\end{eqnarray}
where
\begin{eqnarray}
\rho^2&=&r^2+a^2\cos^2\theta,\quad  f(r)=(r^2+a^2)(1-\frac{r^2}{\ell^2})-2Mr=(1-\frac{a^2}{\ell^2})r^2-2Mr+a^2-\frac{r^4}{\ell^2}\nonumber\\
f(\theta)&=&1+\frac{a^2 \cos^2\theta}{\ell^2},\quad \Sigma=1+\frac{a^2}{\ell^2}.\label{eq2}
\end{eqnarray}
Here $\ell$ is the cosmological radius, $M$ and $a$ are the mass of the black hole and angular momentum per unit mass. The de Sitter space are defined such that  $-\infty\leq t\leq \infty $, $r\geq 0$, $0\leq \theta \leq \pi $, and $0\leq \phi \leq 2\pi$.   The metric (1) describes an interesting rotating AdS black hole called the Kerr-Anti-de Sitter (KAdS) black hole if we replace $\ell^2$ by$-\ell^2$.\\
The only single positive real root is obtained by $r^4-(\ell^2-a^2)+2M\ell^2 r-\ell^2 a^2=0$ and which is located
at the black hole (event) horizon $r_h$ such that
\begin{eqnarray}
r_h&=&\frac{\ell\beta}{\sqrt{3}}.\textrm{sin}\bigg[\frac{1}{3}\textrm{sin}^{-1}\frac{3M\sqrt{3}}{\ell\Sigma\beta}\bigg]\times
\bigg(1+\sqrt{1-\frac{q^2 \ell}{\sqrt{3} M\beta}. \frac{2}{1+\delta}\textrm{cosec}\bigg[\frac{1}{3}
\textrm{sin}^{-1}\frac{3M\sqrt{3}}{\ell\Sigma\beta}\bigg]}\bigg)\label{eq3},
\end{eqnarray}
where
\begin{equation}
\delta=\sqrt{1-\frac{4a^2\beta^2}{3M^2} \textrm{sin}^2\bigg[\frac{1}{3}
\textrm{sin}^{-1}\frac{3\sqrt{3}M}{\ell\Sigma\beta}\bigg]},\quad  \Sigma=1+\frac{a^2}{\ell^2}, \quad \beta=\sqrt{1-\frac{a^2}{\ell^2}}.\label{eq4}
\end{equation}
Expanding  $r_h$ in terms of $\ell$, $M$ and $a$ with $a^2(1+\frac{a^2}{\ell^2})<1$ and setting $\delta=1$, we obtain
\begin{equation}
r_h=\frac{M}{\Sigma}\left(1+\frac{4M^2}{\ell^2{\Sigma\beta^2}}+\cdot\cdot\cdot\right)\left(1+\sqrt{1-\frac{a^2\Sigma}{M^2}}\right)\label{eq5},
\end{equation}
which can be written as
\begin{equation}
r_h=\frac{1}{\Sigma}\left(1+\frac{4M^2}{\ell^2{\Sigma\beta^2}}+\cdot\cdot\cdot\right)\left(M+\sqrt{M^2-a^2\Sigma}\right)\label{eq6},
\end{equation}
It is clear that the event horizon of the Kerr de-Sitter black hole is greater than the Kerr event horizon  $r_{Ke}=M+\sqrt{M^2-a^2}$.  Again it also shows that Kerr black hole \cite{twenty one} horizons for $\ell\rightarrow\infty (\Sigma\longrightarrow1)$ and Schwarzschild-de Sitter black hole \cite{fifty three} horizon for $a=0$.

\section{The Hamilton-Jacobi Method}\label{sec3}

Two new methods have been employed to calculate the imaginary part of the action, one the null geodesic method developed by Parikh and Wilczek \cite{six,seven} and the other method is called Hamilton-Jacobi method \cite{thirty eight, thirty nine, fourty}. The difference of later method from Parikh's is mainly that such method concentrates on introducing the proper spatial distance and upon calculating the relaivistic Hamilton-Jacobi equation. For calculating the imaginary part of the action for the process of s-wave emission across the horizon, which in turn is related, using the WKB approximation \cite{sixty one}, satisfies $\Gamma \sim {\rm exp}(-2{\rm Im}I)$, where $I$ is the action of the outgoing particle and $\Gamma$ is the emission rate.

In the Hamilton-Jacobi method we avoid the exploration of the equation of motion in the Painlev\'e coordinates systems. In order to calculate the imaginary part of the action from the relativistic Hamilton-Jacobi equation, the action $I$ of the outgoing particle from the black hole horizon satisfies the relativistic Hamilton-Jacobi equation
\begin{equation}
g^{ab}\left(\partial_a I\right)\left(\partial_b I\right)+m^2=0,\label{eq7}
\end{equation}
in which $m$ and $g^{ab}$ are the mass of the particle and the inverse metric tensors respectively. We perform the following effective transformation to study the Hawking radiation of the KdS black hole such that
\begin{eqnarray}
\frac{d\phi}{dt}=-\frac{g_{03}}{g_{33}}=\frac{a\Sigma[f(\theta)(r^2+a^2)-f(r)]}{f(\theta)(r^2+a^2)^2-f(r)a^2\sin^2\theta}.\label{eq8}
\end{eqnarray}
Now applying Eq. (\ref{eq8}) on the line element given by Eq. (\ref{eq1}), then the new line element of the Kerr-de Sitter black hole takes on form as
\begin{eqnarray}
ds^2=-\frac{f(r)f(\theta)\rho^2}{f(\theta)(r^2+a^2)^2-f(r)a^2\sin^2\theta}dt^2+\frac{\rho^2}{f(r)}dr^2+\frac{\rho^2}{f(\theta)}d\theta^2.\label{eq9}
\end{eqnarray}
The position of the event horizon is same as given in Eq. (\ref{eq6}). Near the event horizon, the line element given by Eq. (\ref{eq9}) can be written as
\begin{eqnarray}
ds^2=-\frac{f_{,r}(r_h)(r-r_h)\rho^2(r_h)}{(r_h^2+a^2)^2}dt^2+\frac{\rho^2(r_h)}{f_{,r}(r_h)(r-r_h)}dr^2+\frac{\rho^2(r_h)}{f(\theta)}d\theta^2.\label{eq10}
\end{eqnarray}
In which $\rho^2(r_h)=r_h^2+a^2\cos^2\theta$ and $f_{,r}(r_h)=\frac{df}{dr}\bigg |_{r=r_h}=\frac{2}{\Sigma^2}(\beta^2r_h-M-2\frac{r^3_h}{\ell^2})$.
The non-null inverse metric tensors for the metric (\ref{eq10}) are, namely
\begin{eqnarray}
\bar{g}^{11}=-\frac{(r_h^2+a^2)^2}{f_{,r}(r_h)(r-r_h)\rho^2(r_h)}, g^{22}=\frac{f_{,r}(r_h)(r-r_h)}{\rho^2(r_h)}, g^{33}=\frac{f(\theta)}{\rho^2(r_h)}.\label{eq11}
\end{eqnarray}
We can write Eq. (\ref{eq7}) with the help of  (\ref{eq11}) as
\begin{eqnarray}
-\frac{(r^2_h+a^2)^2}{\rho^2(r_h)f_{,r}(r_h)(r-r_h)}\left(\partial_t I\right)^2+\frac{f_{,r}(r_h)(r-r_h)}{\rho^2(r_h)}\left(\partial_r I\right)^2
+\frac{f(\theta)}{\rho^2(r_h)}\left(\partial_\theta I\right)^2+m^2=0\label{eq12}.
\end{eqnarray}
To find the solution of the action $I$ for $I(t, r, \theta, \phi)$ in a easy way, we consider the properties of black hole spacetime, the separation of variables can be taken as follows
\begin{equation}
I=-\omega t+R(r)+H(\theta)+j\phi\label{eq13},
\end{equation}
where $\omega$ is the energy of the particle, $R(r)$ and $H(\theta)$ are the generalized momentums, and $j$ is the angular momentum with respect to $\phi$-axis. So we have $\partial_t I=-\omega+j\Omega_h,$ $\partial_r I=\partial_r R(r)$, $\partial_\theta I=\partial_\theta H$,  where $\Omega_h=\frac{d\phi}{dt}\Bigg |_{r=r_h}=\frac{a\Sigma}{r_h^2+a^2}$ is the angular velocity at the event horizon and $j=(Ma)/\Sigma^2$.

Therefore, inserting  above values into Eq. (\ref{eq12}) and solving $R(r)$ yields an expression of
\begin{eqnarray}
R(r)&=&\pm\frac{r^2_h+a^2}{f_{,r}(r_h)}\int \frac{dr}{(r-r_h)}\times\sqrt{(\omega-j\Omega_h)^2-\frac{\rho^2(r_h)f_{,r}(r_h)(r-r_h)}{(r^2_h+a^2)^2}\left[\frac{f(\theta)}{\rho^2(r_h)}.(\frac{\partial H}{\partial \theta})^2+m^2\right]}\nonumber.
\end{eqnarray}
Finishing the above integral, we get
\begin{eqnarray}
R(r)=\pm \frac{\pi i( r^2_h+a^2)}{f_{,r}(r_h)}(\omega-j\Omega_h)+\xi \label{eq14},
\end{eqnarray}
where $\pm$ sign comes from the square root and $\xi$ is the constant of integration. Inserting Eq. (\ref{eq14}) into Eq. (\ref{eq13}), the imaginary part of two different actions corresponding to the outgoing and incoming particles can be written as
\begin{eqnarray}
{\rm Im}I_\pm =\pm \frac{\pi( r^2_h+a^2)}{f_{,r}(r_h)}(\omega-j\Omega_h)+{\rm Im}(\xi) \label{eq15}.
\end{eqnarray}
By the classical limit \cite{sixty two}, we make ensure the incoming probability to be unity when there is no reflection i.e., everything is absorbed by the horizon. In this case the appropriate value of $\xi$ instead of zero or infinity can be taken as $\xi=\frac{\pi i( r^2_h+a^2)}{f_{,r}(r_h)}(\omega-j\Omega_h) +{\rm Re}(\xi)$. Therefore, ${\rm Im}I_-=0$ and $I_+$ give the imaginary part of action $I$ corresponding to the outgoing particle, namely
\begin{eqnarray}
{\rm Im}I &=& \frac{2\pi(r^2_h+a^2)}{f_{,r}(r_h)}(\omega-j\Omega_h) \nonumber\\
&=&\frac{\Sigma^2\pi( r^2_h +a^2)}{\beta^2r_h-M-2\frac{r^3_h}{\ell^2}}(\omega-j\Omega_h)\label{eq16}.
\end{eqnarray}
Using Eq. (\ref{eq6}) into Eq. (\ref{eq16}), we get the imaginary part of the action as
\begin{eqnarray}
&&{\rm Im}I
=\frac{\pi\left(1+\frac{4M^2}{\ell^2\Sigma\beta^2}+\cdot
\cdot\right)^2{(M+\sqrt{M^2-a^2\Sigma})^2}}{\frac{\beta^2}{\Sigma}\left(1+\frac{4M^2}{\ell^2\Sigma\beta^2}+\cdot
\cdot\right){(M+\sqrt{M^2-a^2\Sigma})}-M-A}\omega\nonumber\\
&&+\frac{\Sigma^2\pi a^2}{\frac{\beta^2}{\Sigma}\left(1+\frac{4M^2}{\ell^2\Sigma\beta^2}+\cdot
\cdot\right){(M+\sqrt{M^2-a^2\Sigma})}-M-A}\omega
-\frac{\Sigma^3\pi a}{\frac{\beta^2}{\Sigma}\left(1+\frac{4M^2}{\ell^2\Sigma\beta^2}+\cdot
\cdot\right){(M+\sqrt{M^2-a^2\Sigma})}-M-A}j,\nonumber
\end{eqnarray}
where $A=\frac{2}{\ell^2\Sigma^3}\left(1+\frac{4M^2}{\ell^2\Sigma\beta^2}+\cdot\cdot\right)^3(M+\sqrt{M^2-a^2\Sigma})^3.$
\begin{eqnarray}
{\rm Im}I
&=&\frac{\pi{(M+\sqrt{M^2-a^2\Sigma})^2}}{\frac{\beta^2}{\Sigma}\left[\left(1-\frac{4M^2}{\ell^2\Sigma\beta^2}+\cdot
\cdot\right){(M+\sqrt{M^2-a^2\Sigma})}-\frac{M\Sigma}{\beta^2}\left(1-\frac{8M^2}{\ell^2\Sigma\beta^2}+\cdot\cdot\right)-B\right]}\omega\nonumber\\
&&+\frac{\Sigma^2\pi a^2}{\frac{\beta^2}{\Sigma}\left[\left(1+\frac{4M^2}{\ell^2\Sigma\beta^2}+\cdot
\cdot\right){(M+\sqrt{M^2-a^2\Sigma})}-\frac{M\Sigma}{\beta^2}-\frac{\Sigma A}{\beta^2}\right]}\omega\nonumber\\
&&-\frac{\Sigma^3\pi a}{\frac{\beta^2}{\Sigma}\left[\left(1+\frac{4M^2}{\ell^2\Sigma\beta^2}+\cdot
\cdot\right){(M+\sqrt{M^2-a^2\Sigma})}-\frac{M\Sigma}{\beta^2}-\frac{\Sigma A}{\beta^2}\right]}j,\nonumber
\end{eqnarray}
where $B=\frac{2}{\ell^2\beta^2\Sigma^2}\left(1+\frac{4M^2}{\ell^2\Sigma\beta^2}+\cdot\cdot\right)(M+\sqrt{M^2-a^2\Sigma})^3.$

Now for the simplicity, neglecting $M^3$ and its higher order terms, we then get
\begin{eqnarray}
{\rm Im}I &=&\frac{\pi\Sigma}{\beta^2}.\frac{(M+\sqrt{M^2-a^2\Sigma})^2}{(M+\sqrt{M^2-a^2\Sigma})-\frac{M\Sigma}{\beta^2}}\omega
+\frac{\Sigma^3\pi a^2}{\beta^2\left[M+\sqrt{M^2-a^2\Sigma}-\frac{M\Sigma}{\beta^2}\right]}\omega\nonumber\\
&&-\frac{\Sigma^4\pi a}{\beta^2\left[M+\sqrt{M^2-a^2\Sigma}-\frac{M\Sigma}{\beta^2}\right]}j.\label{eq17}
\end{eqnarray}
We now focus on a classical treatment of the associated radiation and adopt the picture of a pair of virtual particles spontaneously created just inside the horizon. The positive energy virtual particle can tunnel out while the negative one is absorbed by the black hole resulting in a decrease in the mass. We consider the emitted particle as an ellipsoid shell of energy $\omega$ and  fix the Arnowitt-Deser-Misner(ADM) mass and angular momentum of the total spacetime since in presence of cosmological constant KdS spacetime is dynamic and allow the KdS black hole to fluctuate. When a particle with energy $\omega$ and angular momentum $j$ tunnels out, the mass and angular momentum of the KdS black hole changed into $M-\omega$ and $J-j$. Assuming the self-gravitational interaction into account, the imaginary part of the true action can be calculated from Eq. (\ref{eq17}) in the following integral form
\begin{eqnarray}
{\rm Im}I&=&\frac{\pi\Sigma}{\beta^2}.\int^\omega_0\frac{(M+\sqrt{M^2-a^2\Sigma})^2}{\sqrt{M^2-a^2\Sigma}+(M-\frac{M\Sigma}{\beta^2})}d\omega'+
\frac{\pi\Sigma^3}{\beta^2}.\int^\omega_0\frac{a^2}{\sqrt{M^2-a^2\Sigma}+(M-\frac{M\Sigma}{\beta^2})}d\omega'\nonumber\\
&&-\frac{\pi\Sigma^4}{\beta^2}.\int^j_0\frac{a}{\sqrt{M^2-a^2\Sigma}+(M-\frac{M\Sigma}{\beta^2})}d\omega'\label{eq18}
\end{eqnarray}
For the maximum value of integration, neglecting $(1-\frac{\Sigma}{\beta^2})M$ and, then replacing $M$ by $M-\omega$ and $j$ by $J-j$, equation (\ref{eq18}) becomes
\begin{eqnarray}
{\rm Im}I&=&-\frac{\pi\Sigma}{\beta^2}.\int^{(M-\omega)}_M\frac{(M-\omega+\sqrt{(M-\omega)^2-a^2\Sigma})^2}{\sqrt{(M-\omega)^2-a^2\Sigma}}d(M-\omega')\nonumber\\
&&-\frac{\pi\Sigma^3}{\beta^2}.\int^{(M-\omega)}_M\frac{a^2}{\sqrt{(M-\omega)^2-a^2\Sigma}}d(M-\omega')\nonumber\\
&&+\frac{\pi\Sigma^4}{\beta^2}.\int^{(J-j)}_J\frac{a}{\sqrt{(M-\omega)^2-a^2\Sigma}}d(J-j'),\label{eq19}
\end{eqnarray}
where $J-j'=(M-\omega')a/\Xi^2$. Therefore Eq. (\ref{eq19}) becomes
\begin{eqnarray}
{\rm Im}I&=&-\frac{\pi\Sigma}{\beta^2}.\int^{(M-\omega)}_M\frac{(M-\omega+\sqrt{(M-\omega)^2-a^2\Sigma})^2}{\sqrt{(M-\omega)^2-a^2\Sigma}}d(M-\omega')\nonumber\\
&=&-\frac{\pi\Sigma}{\beta^2}.\int^{(M-\omega)}_M\frac{2(M-\omega)^2+2(M-\omega)\sqrt{(M-\omega)^2-a^2\Sigma}}{\sqrt{(M-\omega)^2-a^2\Sigma}}d(M-\omega')\nonumber\\
&&+\frac{\pi\Sigma}{\beta^2}.\int^{(M-\omega)}_M\frac{a^2\Sigma}{\sqrt{(M-\omega)^2-a^2\Sigma}}d(M-\omega').\label{eq20}
\end{eqnarray}
Doing the $\omega'$ integral finally yields
\begin{eqnarray}
{\rm Im}I&=&-\frac{\pi\Sigma}{\beta^2}\Big\{(M-\omega)\sqrt{(M-\omega)^2-a^2\Sigma}
+(M-\omega)^2-M\sqrt{M^2-a^2\Sigma}-M^2\Big\}\nonumber\\
&=&-\frac{\pi\Sigma}{2\beta^2}\Big\{2(M-\omega)\sqrt{(M-\omega)^2-a^2\Sigma}
+2(M-\omega)^2-2M\sqrt{M^2-a^2\Sigma}-2M^2\Big\}\nonumber\\
&=&-\frac{1}{2}{\rm exp}[\pi(r^2_f-r^2_i)]\nonumber\\
&=&-\frac{1}{2}{\rm exp}(\Delta S_{BH}),\label{eq21}
\end{eqnarray}
where the Bekenstein-Hawking entropy of the black hole is $S_{BH}(M)=\pi r^2_i$ and $S_{BH}(M-\omega)=\pi r^2_f$ and $\Delta S_{BH}=S_{BH}(M-\omega)-S_{BH}(M)$ is the change of Bekenstein-Hawking entropies of the Kerr-de Sitter black hole before and after the emission of particles by setting $r_f=\frac{\sqrt\Sigma}{\beta}[(M-\omega)+\sqrt{(M-\omega)^2-a^2\Sigma}]$  and  $r_i=\frac{\sqrt\Sigma}{\beta}[M+\sqrt{M^2-a^2\Sigma}]$ respectively.

Utilizing WKB approximation \cite{sixty one} and the relationship between the tunneling rate and the imaginary part of the action of the radiative particle, the tunneling rate for KdS black hole is given by
\begin{eqnarray}
\Gamma \sim {\rm exp}(-2{\rm Im}I)={\rm exp}(\Delta S_{BH}),\label{eq22}
\end{eqnarray}

 \section{Purely thermal radiation}\label{sec4}

The radiation spectrum is not pure thermal although gives a correction to the Hawking radiation of KdS black hole as point out by  Eq. (\ref{eq22}). In the form of a thermal spectrum, using the WKB approximation the tunneling rate is also related to the energy and the Hawking temperature of the radiative particle as
$\Gamma \sim {\rm exp}(-\frac{\Delta \omega}{T})$. If $\Delta \omega < 0$ is the energy of the emitted particle then due to energy conservation, the energy of the outgoing shell must be $-\Delta \omega$, then above expression becomes
\begin{eqnarray*}
\Gamma \sim {\rm exp}(\frac{\Delta \omega}{T})
\end{eqnarray*}
Now using the first law of thermodynamics $(\Delta \omega=T\Delta S$), we can write  $\Gamma \sim {\rm exp}({\Delta S})$, which is related to the change of Bekenstein-Hawking entropy as follows
\begin{eqnarray}
\Gamma \sim {\rm exp}({\Delta S_{BH}})={\rm exp}\{S_{BH}(M-\omega)-S_{BH}(M)\}\label{eq23}
\end{eqnarray}
We establish Eq.(\ref{eq23}) as developed by Rahman et al. \cite{fifty three} in power of $\omega$ upto second order using Taylor\rq s theorem of the form
\begin{eqnarray}
\Gamma \sim {\rm exp}(\Delta S_{BH})={\rm exp}\left\{-\omega \frac{\partial S_{BH}(M)}{\partial M}+\frac{\omega^2}{2}\frac{\partial^2 S_{BH}(M)}{\partial M^2}\right\} ,\label{eq24}
\end{eqnarray}
From Eq.(\ref{eq21}), we can write
\begin{eqnarray}
 S_{BH}(M-\omega)=\frac{\pi\Sigma}{\beta^2}[(M-\omega)+\sqrt{(M-\omega)^2-a^2\Sigma}]^2 .\label{eq25}
 \end{eqnarray}
At $\omega = 0$,
\begin{eqnarray}
\frac{\partial S_{BH}(M)}{\partial M}=\frac{2\Sigma}{\beta^2}\left[2M+\sqrt{M^2-a^2\Sigma}+\frac{M^2}{\sqrt{M^2-a^2\Sigma}}\right]\label{eq26}
\end{eqnarray}
\begin{eqnarray}
\frac{\partial^2 S_{BH}(M)}{\partial M^2}=\frac{2\Sigma}{\beta^2}\left[2+\frac{3M}{\sqrt{M^2-a^2\Sigma}}-\frac{M^3}{(M^2-a^2\Sigma)^{\frac{3}{2}}}
\right]\label{eq27}
\end{eqnarray}

Therefore, the tunneling rate in power of $\omega$ upto second order, the purely thermal spectrum can be revealed from Eq.(\ref{eq24}) as follows:
\begin{eqnarray}
\Gamma \sim {\rm exp}(\Delta S_{BH})={\rm exp}[\pi (-\omega\gamma+\frac{\omega^2}{2}\lambda)],\label{eq28}
\end{eqnarray}

where $\gamma=\frac{2\Sigma}{\beta^2}\left[2M+\sqrt{M^2-a^2\Sigma}+\frac{M^2}{\sqrt{M^2-a^2\Sigma}}
\right]$ and  $\lambda=\frac{2\Sigma}{\beta^2}\left[2+\frac{3M}{\sqrt{M^2-a^2\Sigma}}-\frac{M^3}{(M^2-a^2\Sigma)^{\frac{3}{2}}}
\right].$ \\

The radiation spectrum given by  (\ref{eq28}) is more accurate and provides an interesting correction to Hawking pure thermal spectrum.

\section{Concluding Remarks}\label{sec5}

In this paper, we have discussed the purely thermal and non-thermal Hawking radiations as massive particle tunneling method from KdS black hole by taking into account the self-gravitational interaction, the background spacetime is dynamical and the energy as conservation. We have explored that the tunneling rate at the event horizon of KdS black hole is related to the Bekenstein-Hawking entropy. In the limiting case $\Sigma=1, \beta=1$ the KdS black hole is reduced to Kerr black hole \cite{twenty one} . The positions of the event horizon of Kerr black hole before and after the emission of the particles with energy $\omega$ are $r_i=M+\sqrt{M^2-a^2}$ and $r_f=(M-\omega)+\sqrt{(M-\omega)^2-a^2}$. From Eq. (\ref{eq22}), the non-thermal radiation for Kerr black hole becomes
\begin{eqnarray}
\Gamma \sim {\rm exp}(-2{\rm Im}I)&=&{\rm exp}\left\{\pi\left[\left\{(M-\omega)+\sqrt{(M-\omega)^2-a^2}\right\}^2-\left\{M+\sqrt{M^2-a^2}\right\}^2\right]\right\}\nonumber\\
&=&{\rm exp}[\pi(r^2_f-r^2_i)]={\rm exp}(\Delta S_{BH}).\label{eq29}
\end{eqnarray}
and the purely thermal radiation becomes
\begin{eqnarray}
&&\Gamma \sim {\rm exp}(\Delta S_{BH})\nonumber\\
&&={\rm exp}\Bigg[-2\pi\omega\Big\{\Big(2M+\sqrt{M^2-a^2}+\frac{M^2}{\sqrt{M^2-a^2}}\Big)\nonumber\\
&&-\frac{\omega}{2}\Big(2+\frac{3M}{\sqrt{M^2-a^2}}-\frac{M^3}{(M^2-a^2)^{\frac{3}{2}}}\Big)\Big\}\Bigg],\label{eq30}
\end{eqnarray}
  It is also clear that both the results are fully consistent with that obtained by Parikh and Wilczek \cite{six} and also reduced to our previous result of SdS \cite{fifty three} black hole when $a=0$.

In addition, our discussion made here can be directly to the anti-de Sitter case by changing the sign of the cosmological constant to a negative one, and also can be easily generalized to the other black holes.\\

This research is supported by the Abdus Salam International Centre for Theoretical Physics (ICTP), Trieste, Italy.

\end{document}